\documentclass[aps,prl,superscriptaddress,
               twocolumn,balancelastpage,
               longbibliography,
               ]{revtex4-2}

\usepackage[colorlinks,bookmarks=false,citecolor=blue,linkcolor=blue,urlcolor=blue]{hyperref}
\usepackage[all]{hypcap}   

\usepackage{amsmath,amssymb}
\usepackage{graphicx}
\graphicspath{{Figures/}{/}}

\usepackage{verbatim}
\usepackage{color}

\usepackage{placeins}  
\usepackage{flafter}     

\usepackage{color}

\makeatletter
\let\Hy@backout\@gobble
\makeatother

\begin{document}

\title{
Resonance states at Casati wave numbers for the 3-disk billiard 
}

\author{Roland Ketzmerick}
\affiliation{Technische Universit\"at Dresden,
    Institut f\"ur Theoretische Physik and Center for Dynamics,
    01062 Dresden, Germany}

\author{Jan Robert Schmidt}
\affiliation{Technische Universit\"at Dresden,
    Institut f\"ur Theoretische Physik and Center for Dynamics,
    01062 Dresden, Germany}

\date{\today}
\begin{abstract}
Resonance states of the 3-disk scattering system are presented for the 
first Casati wave number $k \approx 912$
and the second Casati wave number $k \approx 91242$.
They show multifractal structure in phase space, similar
to the pioneering work by Casati \emph{et al.} 
[Physica D \textbf{131}, 311 (1999)] 
for an open chaotic quantum map.
In position space we observe scarring along segments of rays,
related to multifractality and universal fluctuations,
as recently found for dielectric cavities.
To the best of our knowledge 
this resonance state at the second Casati wave number
has a much larger wave number than
published resonance states for the 3-disk scattering system
or any other open or closed chaotic billiard.

\end{abstract}

\maketitle

The multifractal structure of resonance states in scattering systems with classically chaotic dynamics was first 
presented by Casati, Maspero, and Shepelyansky~\cite{CasMasShe1999b, CasMasShe1999:p}
for the quantum kicked rotor with absorption. 
It was found that resonance states are supported by the 
backward trapped set of the classical
scattering system, which is a fractal.
The multifractal structure on the backward trapped set
depends systematically on the decay rate of the resonance states,
which was recently analyzed for dielectric cavities~\cite{KetClaFriBae2022}. 

This is a brief report on the occasion of the 80th birthday of Giulio Casati,
born on {\bf 9.12.42}, see also \cite{Rob2022}.
We present resonance states of the 3-disk scattering system
~\cite{GasRic1989c, Wir1999, WeiBarKuhPolSch2014},
a paradigmatic model of chaotic scattering.

\begin{figure}[h!]
	\includegraphics[scale=1.05]{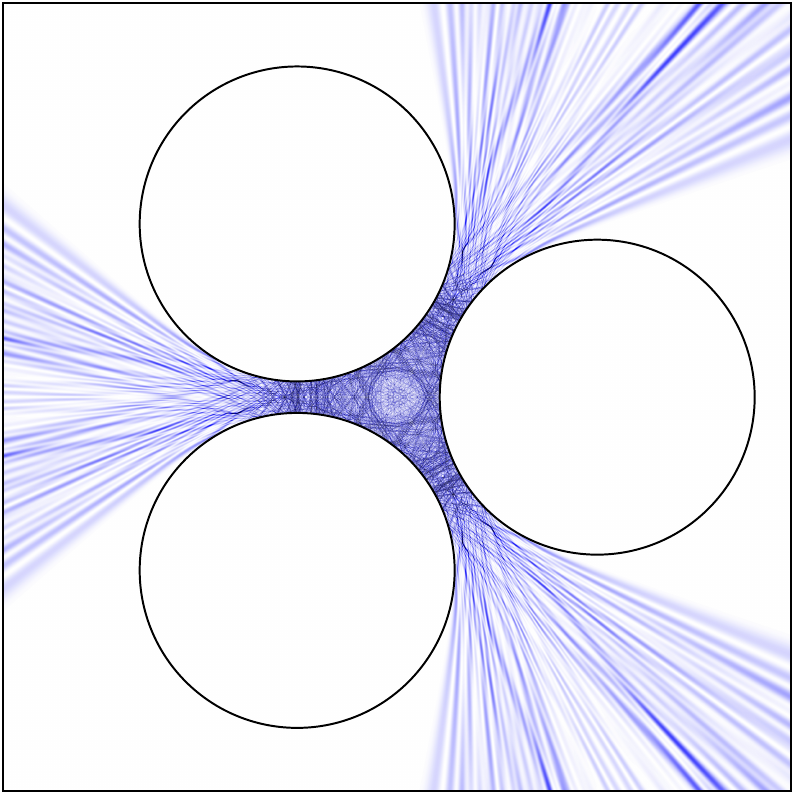}
	\caption{
		Intensity of a resonance state at the first Casati wave number
		$k = {\bf 912}.4484 - 0.3112\,$i.
	}
	\label{FIG:position_912}
\end{figure}

A resonance state
at the first Casati wave number $k \approx {\bf 912}$
is shown in Fig.~\ref{FIG:position_912}.
We consider a
distance $R=2.2$ between the centers of the disks, while
the radius of the disks is set to one.
One can see the location of the three disks
and the 6-fold symmetry of the intensity of the resonance state.

\begin{figure*}[t!]
	\includegraphics[scale=1.12]{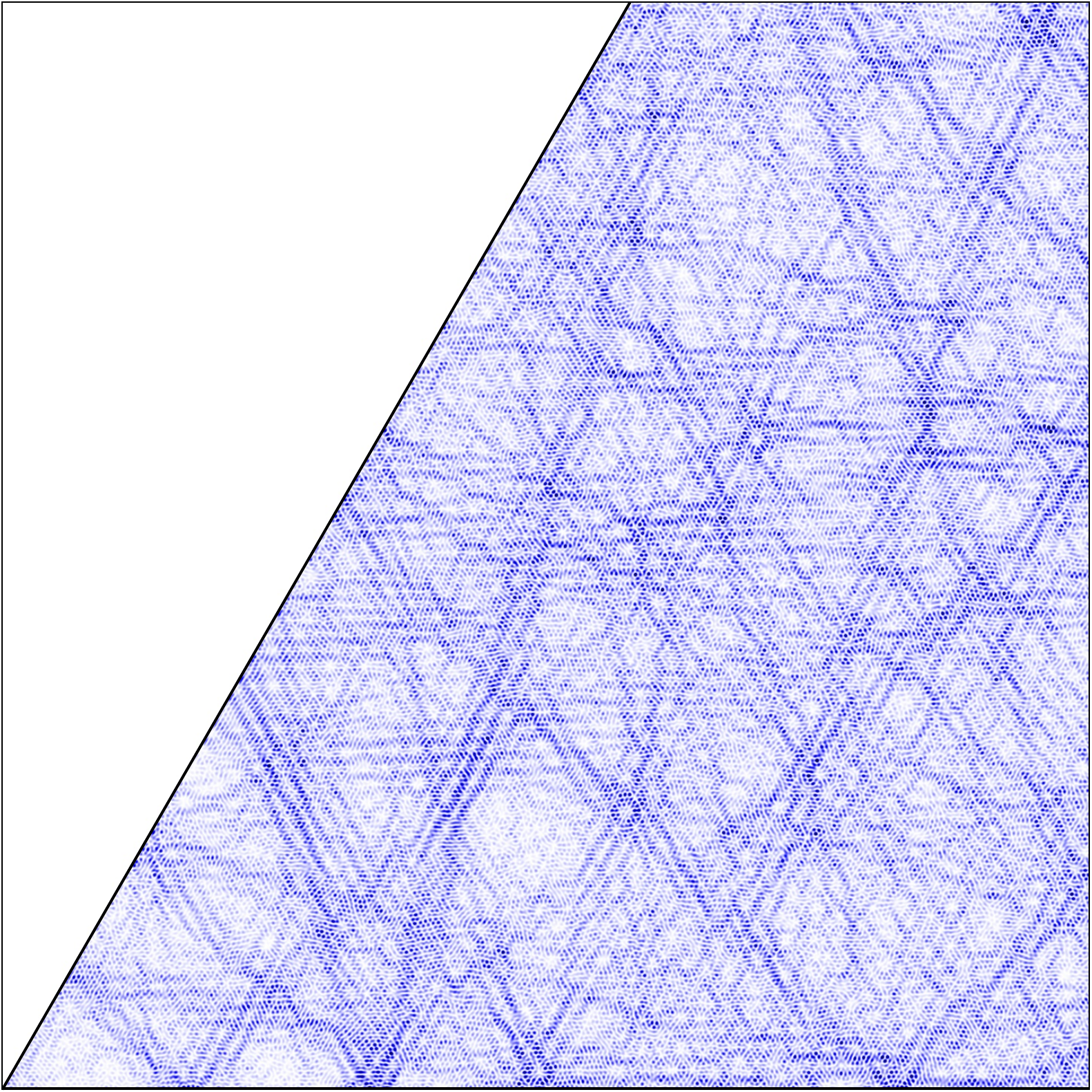}
	\caption{
		Intensity of a resonance state at the second Casati wave number
		$k = {\bf 91242}.2841 - 0.6951\,$i
		for a symmetry region near the origin $[0,0.01] \times [0,0.01]$,
		showing ray-segment scarring.
		\vspace*{1.cm}
	}
	\label{FIG:position_91242}
\end{figure*}

\begin{figure*}[t!]
	\includegraphics[scale=1.12]{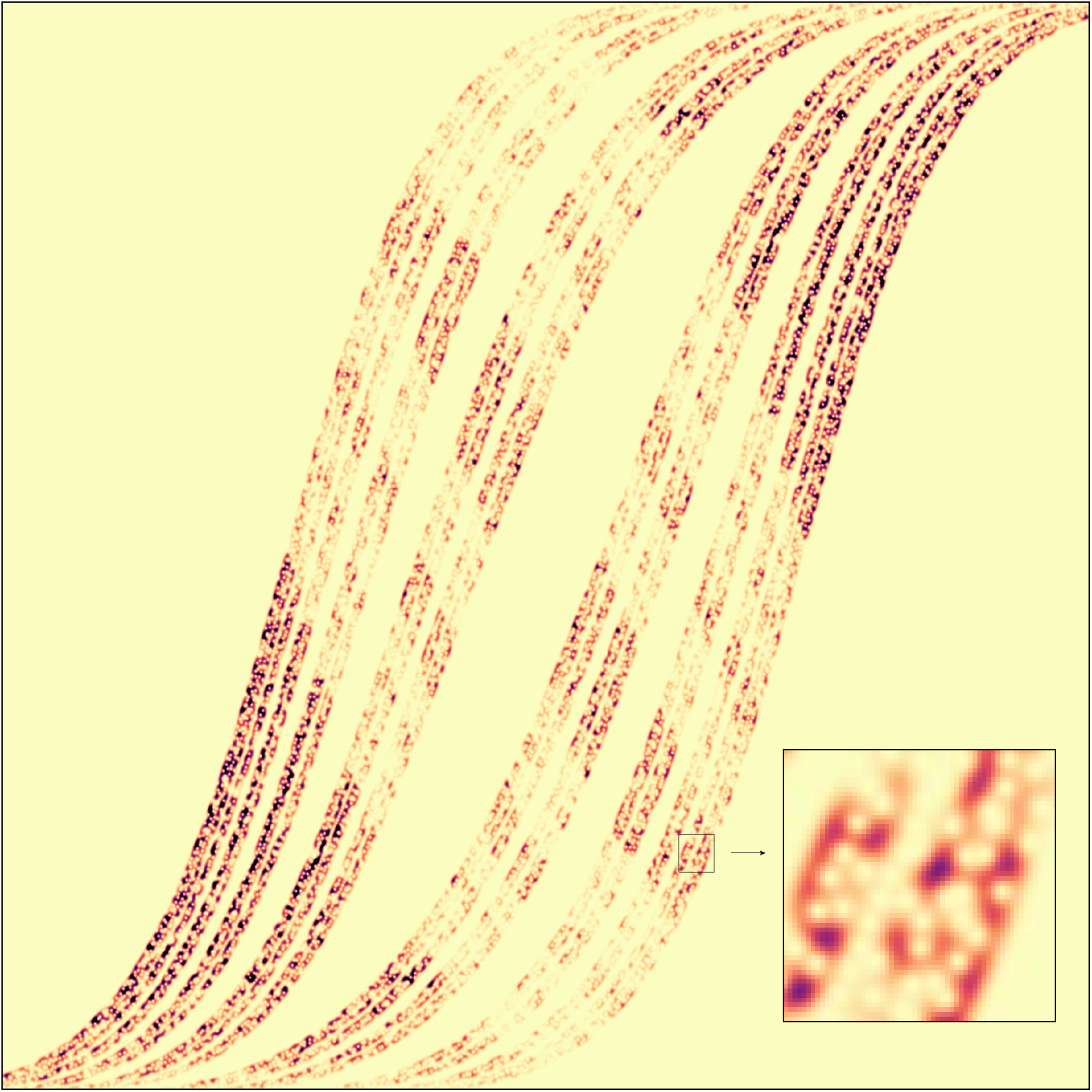}
	\caption{
		Boundary Husimi function of a resonance state at the 
		second Casati wave number
		$k = {\bf 91242}.2841 - 0.6951\,$i
		for Birkhoff coordinates $s \in [-1.1, 1.1]$ and $p \in [-1, 1]$,
		showing a multifractal structure on fine scales. 
		A magnification shows that the Husimi function is smooth on the scale of a Planck cell.
		\vspace*{1.cm}
	}
	\label{FIG:husimi_91242}
\end{figure*}

At the second Casati wave number $k \approx {\bf 91242}$
the intensity of a resonance state is shown in a symmetry region
near the origin,
see Fig.~\ref{FIG:position_91242}.
One can observe scarring along many ray segments.
Often it continues beyond a reflection on the boundary.
This is in agreement with recent findings for resonance modes in 
dielectric cavities~\cite{KetClaFriBae2022}.
There it was argued, that this scarring is
based on multifractality of resonances in phase space
and universal fluctuations.
It conceptually differs from periodic-orbit scarring and
becomes even more prominent in the semiclassical limit.

To the best of our knowledge 
this resonance state at the second Casati wave number
has a much larger wave number than
published resonance states for the 3-disk scattering system
or any other open or closed chaotic billiard.
We implemented the matrix $M$~\cite{GasRic1989c}, where
$\det (M) = 0$ gives the poles of the scattering matrix,
by a fast matrix-vector multiplication~\cite{KetSch2023:p}.

The multifractal structure in phase space on the boundary of a disk
is shown with unprecedented resolution
in Fig.~\ref{FIG:husimi_91242}.
One observes that the resonance state is supported by the backward trapped 
set of the chaotic saddle. 
This is well understood since the work by 
Casati \emph{et al.}~\cite{CasMasShe1999b}.
Additionally, one observes a multifractal structure on the
backward trapped set. This structure strongly depends on the
decay rate of the resonance. Here we show just one resonance state
at an exemplary decay rate related to the imaginary part of $k$.
There exist heuristic approximations for this structure in
fully open quantum maps
as well as in partially open quantum maps
and dielectric cavities, see references in~\cite{KetClaFriBae2022}.
A semiclassical theory for the multifractal structure of resonance states
in chaotic scattering systems,
however, is still missing more than 20 years after the 
pioneering work of 
Casati \emph{et al.}~\cite{CasMasShe1999b}.

\vspace*{0.2cm}

RK thanks Giulio Casati for more than three decades of stimulating
scientific discussions, his support in many ways, and his 
valuable and long-lasting contributions to the quantum chaos 
community.


%

\end{document}